\documentclass{ws-ijmpe}

\usepackage[super,compress]{cite}
\usepackage[dvips]{hyperref}
\hypersetup{colorlinks,urlcolor=black,citecolor=black,linkcolor=black,filecolor=black}
\usepackage{breakurl}

\usepackage{hyperref}
\usepackage{pstricks}
\usepackage{mathrsfs}

\def\apj{Astrophys. J.}

\def\actaa{Acta Astronomica}

\def\prd{Phys. Rev. D}
\def\prc{Phys. Rev. C}
\def\pasj{PASJ}

\begin{document}

\markboth{Kuantay Boshkayev, Jorge Rueda and Remo Ruffini}{On the maximum mass of general relativistic uniformly rotating white dwarfs}

\catchline{}{}{}{}{}

\title{ON THE MAXIMUM MASS OF GENERAL RELATIVISTIC UNIFORMLY ROTATING WHITE DWARFS 
}

\author{\footnotesize KUANTAY BOSHKAYEV, JORGE RUEDA AND REMO RUFFINI
}

\address{Dipartimento di Fisica, Universit\`a di Roma "La Sapienza",\\ Piazzale Aldo Moro 5,
I-00185 Roma, Italy\\
ICRANet, Piazzale della Repubblica 10, I-65122 Pescara, Italy.
\\
kuantay@icra.it, jorge.rueda@icra.it and ruffini@icra.it}

%

\maketitle


\begin{abstract}
The properties of uniformly rotating white dwarfs are analyzed within the framework of general relativity. Hartle's formalism is applied to construct self-consistently the internal and external solutions to the Einstein equations. The mass, the radius, the moment of inertia and quadrupole moment of rotating white dwarfs have been calculated as a function of both the central density and rotation period of the star. The maximum mass of rotating white dwarfs for stable configurations has been obtained.
\end{abstract}

\section*{}

Equilibrium configurations of non-rotating (static) $^{4}$He, $^{12}$C, $^{16}$O and $^{56}$Fe white dwarfs (WDs) within general relativity (GR) have been recently constructed \cite{RotD2011}. The white dwarf matter has been there described by the relativistic generalization of the Feynman-Metropolis-Teller (RFMT) equation of state (EoS) obtained by Rotondo et al.\cite{RotC2011}. A new mass-radius relation that generalizes both the works of Chandrasekhar\cite{chandrasekhar31} and Hamada \& Salpeter\cite{hamada61} has been there obtained, leading to a smaller maximum mass and a larger minimum radius with respect to the previous calculations. In addition, it has been shown how both GR and inverse $\beta$-decay are relevant for the determination of the maximum stable mass of non-rotating WDs. 

It is therefore of interest to generalize the above results to the case of rotation. As a first attempt, we construct in this article general relativistic uniformly rotating WDs in the simplified case when microscopic Coulomb screening is neglected in the EoS, i.e. we follow the Chandrasekhar approximation by describing the matter as a locally uniform fluid of electrons and a lattice of nuclei\cite{chandrasekhar31}. In this case the EoS depends only on the density and the average molecular weight $\mu=A/Z$ ($A$ is the mass number and $Z$ is the number of protons in a nucleus) which, as usual, we fix to the value $\mu=2$.

We apply Hartle's formalism to the description of the structure of rotating objects, up to second order terms in the angular velocity of the star $\Omega$. In this ``slow'' approximation regime, the solution of the Einstein equations in the exterior can be written in analytic closed form in terms of the mass $M$, angular momentum $J$ and quadrupole moment $Q$ of the star, by perturbing a seed static solution. The corresponding interior solution, which matches with the exterior one, can be then constructed by solving numerically a system of ordinary differential equations. The spacetime geometry, with an appropriate choice of coordinates is, in geometrical units $c=G=1$, described by \cite{H1967,HT1968}
\begin{eqnarray}\label{eq:h1}
ds^2 &=& e^{\nu(r)}\left[1+2h(r,\theta)\right]dt^{2}-e^{\lambda(r)}\left[1+\frac{2m(r,\theta)}{r-M(r)}\right]dr^2 \nonumber \\
&-&r^2\left[1+2k(r,\theta)\right]\left[d\theta^2+\sin^2\theta(d\phi-\omega dt)^2\right]+O(\Omega^3),
\end{eqnarray}
where $e^{\lambda(r)}=[1-2M(r)/r]^{-1}$, $h(r,\theta)=h_0(r)+h_2(r)P_2(\cos\theta)+...$, $m(r,\theta)=m_0(r)+m_2(r)P_2(\cos\theta)+...$, and $k(r,\theta)=k_2(r)P_2(\cos\theta)+...$,
Here $P_2(\cos\theta)$ is the Legendre polynomial of second order, $e^{\nu(r)}$ and $e^{\lambda(r)}$ are the metric functions of the static seed solution, $M(r)$ is the corresponding mass profile of the non-rotating star. The angular velocity of local inertial frames $\omega=\omega(r)$, which is proportional to $\Omega$, as well as the functions $h_0$, $h_2$, $m_0$, $m_2$, $k_2$, proportional to $\Omega^2$, must be calculated from the Einstein equations.

We focus here mainly on the determination of the maximum stable mass of uniformly rotating WDs. The mass of a rotating WD is limited by the so-called mass-shedding instability. If the velocity of a particle on the surface of the star exceeds the velocity of a free-particle in the co-rotating circular orbit, the star starts loosing its mass, becoming thus unstable\cite{stergioulas}. A procedure to obtain the maximum possible angular velocity of the star before reaching this limit was developed in \cite{Friedman1986}. However, in practice, it is less complicated to compute the mass-shedding angular velocity of a star $\Omega_{ms}$, from the orbital angular velocity $\Omega_{orb}$ of a co-rotating test particle in the external field at the equatorial plane. For the Hartle-Thorne external solution, the orbital angular velocity $\Omega_{orb}$ for co-rotating particles can be obtained as\cite{2008AcA....58....1T,Bini2011}
\begin{equation}
\Omega_{orb}(r)=\Omega_{0}(r)\left[1- j F_{1}(r)+j^2F_{2}(r)+qF_{3}(r)\right],
\end{equation}
where $j=c J/(G M^2)$ and $q=c^4 Q/(G^2 M^3)$ are the dimensionless angular momentum and quadrupole moment. The other quantities are defined as follows
\begin{eqnarray*}
&&\Omega_{0}=\frac{ M^{1/2}}{r^{3/2}}, \qquad F_{1}=\frac{M^{3/2}}{r^{3/2}},\\
&&F_{2}=(48{M}^7-80{M}^6r+4{M}^5r^2-18{M}^4r^3+40{M}^3r^4 \\
&&+10{M}^2r^5+15{M}r^6-15r^7)/[16{M}^2r^4(r-2{M})]+F, \\
&&F_{3}=\frac{6{M}^4-8{M}^3r-2{M}^2r^2-3{M}r^3+3r^4}{16{M}^2r(r-2{M})/5}-F,\\
&&F=\frac{15(r^3-2{M^3})}{32{M}^3}\ln\frac{r}{r-2{M}}.
\end{eqnarray*}

The parameters $M$, $J$ and $Q$, are obtained for a given EoS, from the matching procedure between the internal and external solutions. Note, that the total mass is defined by $M=M^{J\neq0}=M^{J=0}+\delta M$, where $M^{J=0}$ is the mass of a static white dwarf with the same central density as $M^{J\neq0}$, and $\delta M$ is the contribution to the mass due to rotation. The value of $\Omega_{ms}$ can be computed by gradually increasing the value of $\Omega$ until it reaches $\Omega_{orb}$. Clearly, the matching is carried out at the surface of the rotating star, i.e. therefore one should set $r=R_{eq}$.

\begin{figure*}
\centering
\begin{tabular}{lr}
\includegraphics[width=6.15 cm,clip]{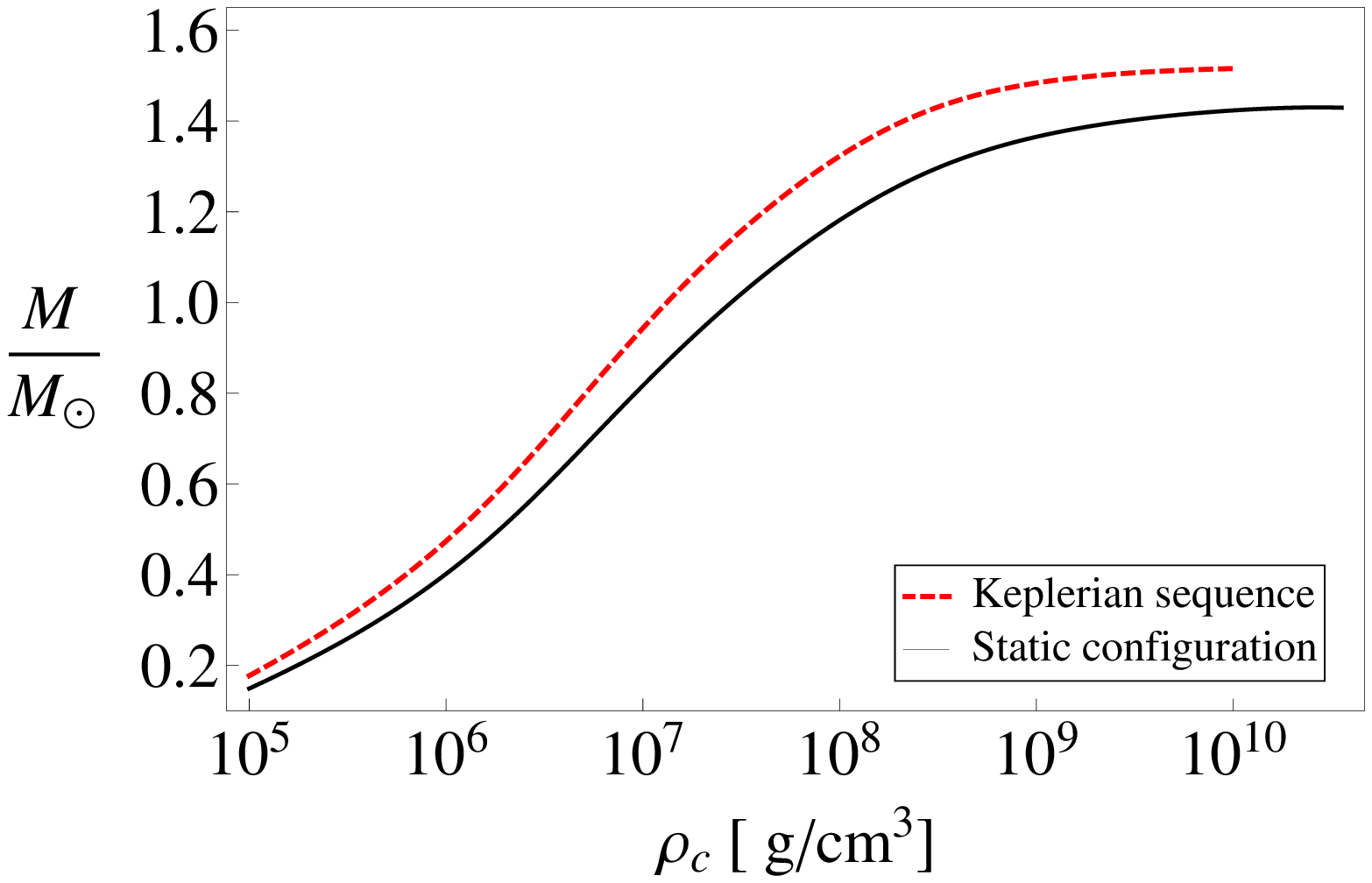} & \includegraphics[width=6.15 cm,clip]{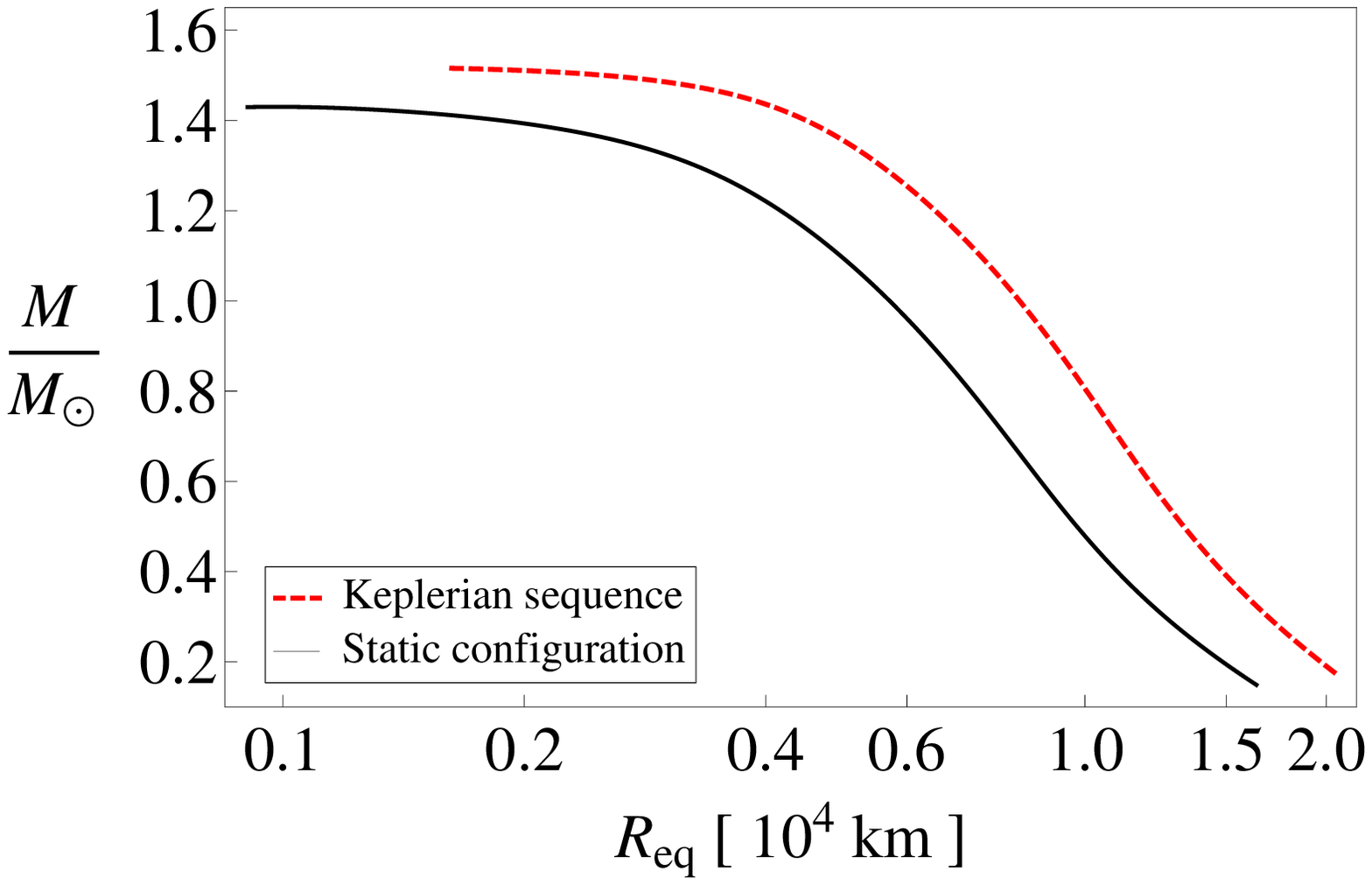}
\end{tabular}
\caption{Left panel: mass in solar masses versus the central density $\rho_c$. Right panel: mass in solar masses versus the equatorial radius $R_{eq}$. The solid curve corresponds to non-rotating WDs and the dashed curve corresponds to the mass, along the Keplerian sequence.}\label{fig:MrhoRCh}
\end{figure*}

\begin{figure*}
\centering
\begin{tabular}{lr}
\includegraphics[width=6 cm,clip]{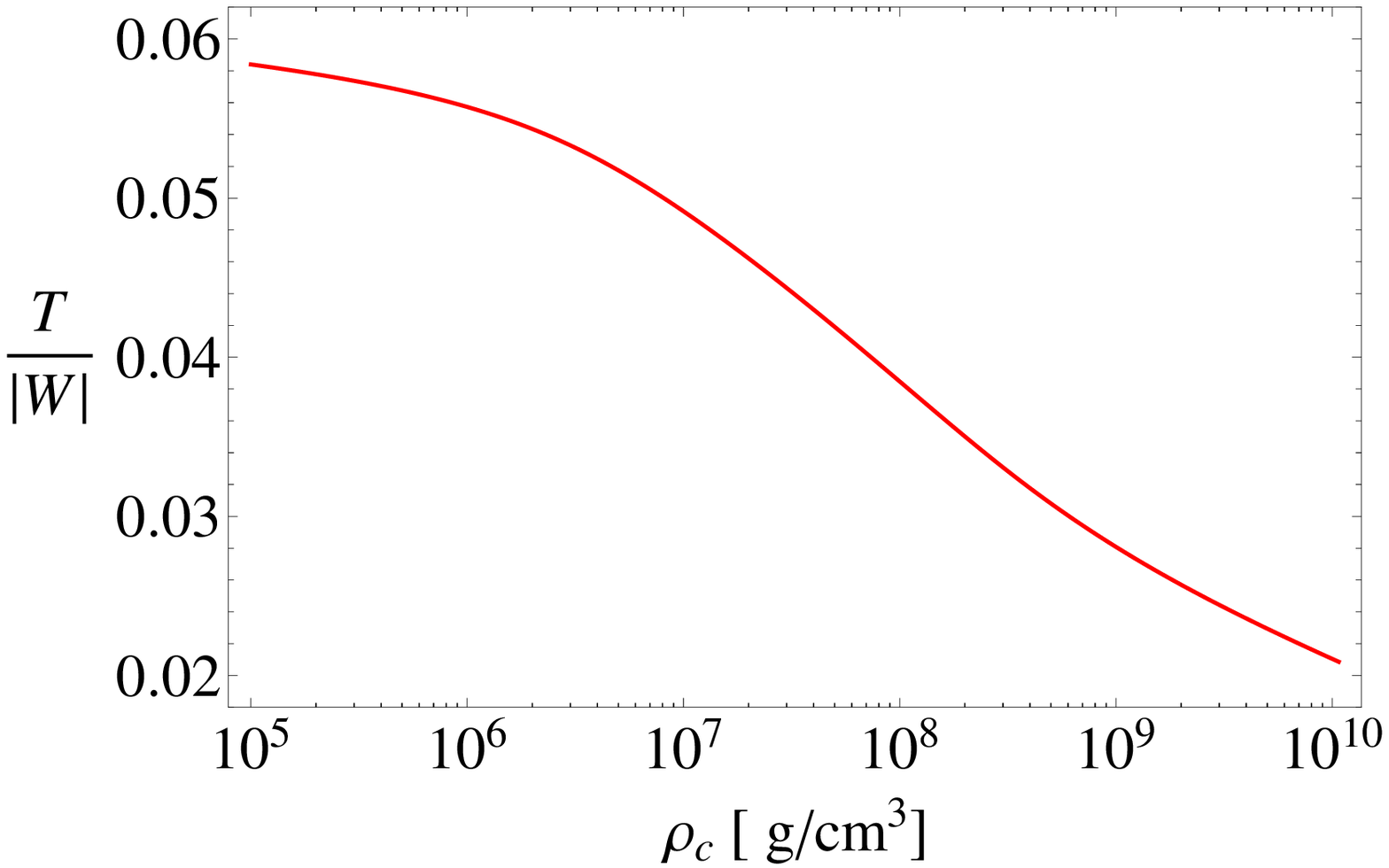} & \includegraphics[width=6 cm,clip]{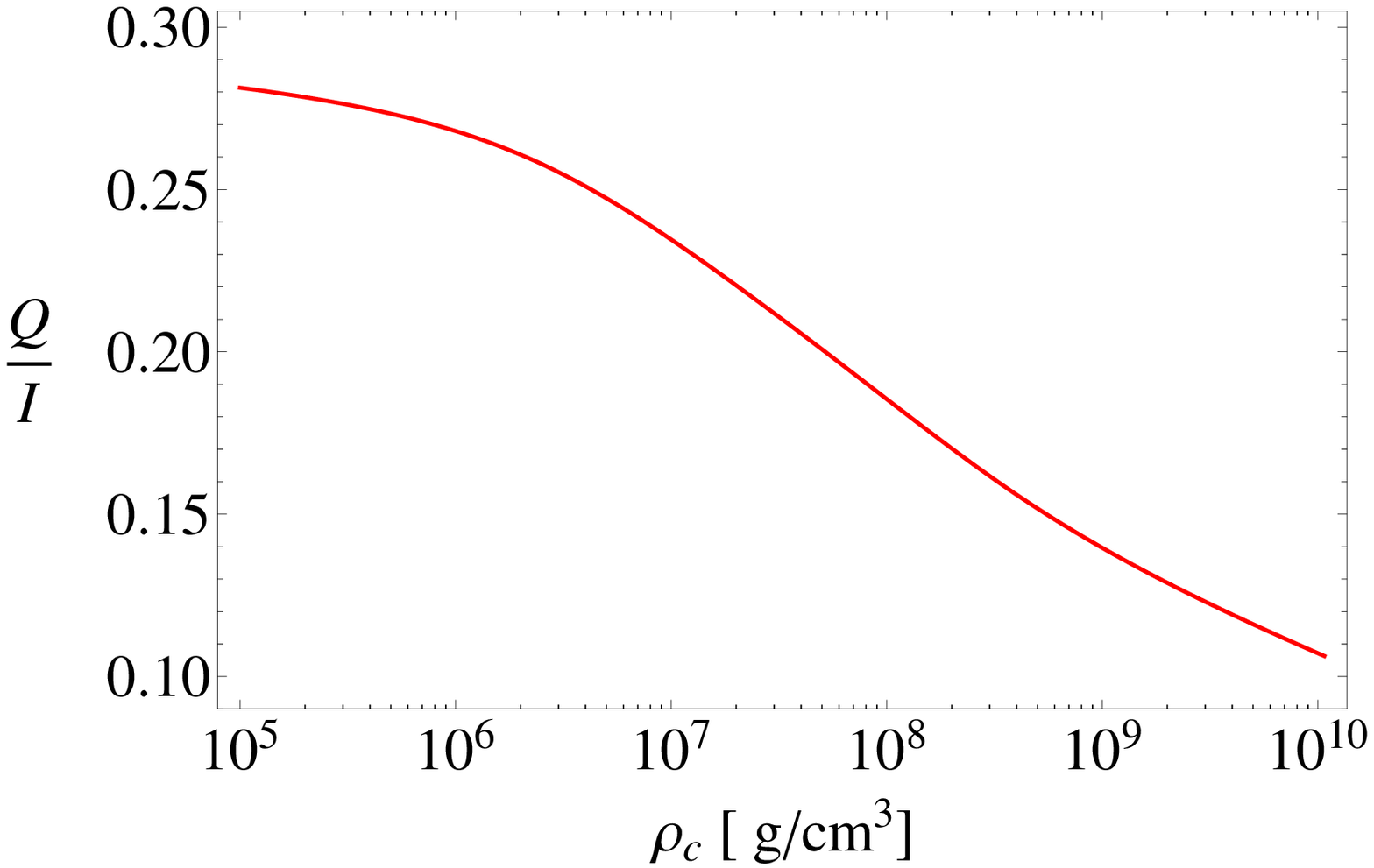}
\end{tabular}
\caption{Left panel: $T/|W|$ is the rotational$/$gravitational energy  versus  $\rho_c$. Right panel: $Q/I$ is the quadrupole moment$/$moment of inertia versus  $\rho_c$, along the Keplerian sequence.}\label{fig:TWQIrhoCh}
\end{figure*}

In Fig.~\ref{fig:MrhoRCh} we show the mass-central density relation (left panel) and the mass-radius relation (right panel) of the Keplerian (mass-shedding) sequence of general relativistic uniformly rotating WDs. In Fig.~\ref{fig:TWQIrhoCh} the ratio between the rotational energy and gravitational energy versus the central density (left panel) and the ratio between the quadrupole moment and moment of inertia versus the central density are shown. With the increasing central density these ratios start to decrease, which shows that the system is gravitationally bound.
%

For the average nuclear composition $\mu=2$, the maximum rotating mass is $M_{max}=1.51595 M_\odot$, and the corresponding polar and equatorial radii are $R_p=1198.91$ km and $R_{eq}=1583.47$ km. For the sake of comparison, for the same EoS, the maximum non-rotating mass is $1.42936 M_\odot$ and the corresponding radius is $R=904.083$ km\cite{RotD2011}. It is interesting to compare the above results with the classic work of Roxburgh and Durney\cite{1966ZA.....64..504R}. They found a maximum mass $1.4825 M_\odot$ for the mass-shedding WD sequence, as well as the critical polar radius 363 km for rotating instability, for the same EoS we used here. The Roxburgh critical radius is rather small with respect to our critical polar radii. It is clear that such a small radius would lead to a configuration with central density over the limit established by inverse $\beta$-decay: the average density obtained for the Roxburgh's critical configuration is $\sim 1.46\times 10^{10}$ g/cm$^3$, very close to $\rho^{\beta}_{\rm crit}=3.97\times 10^{10}$ g/cm$^3$ for $^{12}$C WDs and even closer to $\rho^{\beta}_{\rm crit}=1.94\times 10^{10}$ g/cm$^3$ for $^{16}$O WDs\cite{RotD2011}.

\emph{Discussions and Conclusion}--We have here computed general relativistic configurations of uniformly rotating WDs within the Chandrasekhar approximation for the EoS. We have, in particular, determined the maximum stable mass $M_{max}=1.51595 M_\odot$ for the average nuclear composition $\mu=2$. A realistic description of a WD must take into account the specific chemical composition, the nucleus-electron and electron-electron Coulomb interactions, as well as the strong interactions and the electroweak equilibrium at high densities, all within a self-consistent relativistic framework, as recently done in the case of non-rotating WDs\cite{RotD2011}.

Besides the mass-shedding, there are additional instabilities that might play a fundamental role in the determination of the entire region of stability of rotating WDs, e.g. inverse $\beta$-decay of the composing nuclei and axisymmetric instability\cite{1968ApJ...151.1089O}. This important issue will be addressed in forthcoming publications.

The results presented here open the way to a more general description of a rotating WD and are relevant both for the theory of delayed type Ia supernova explosions as well as for the white dwarf model of Soft Gamma-Ray Repeaters and Anomalous X-Ray Pulsars\cite{M2012}.


\end{document}